\documentclass{moriond}

\usepackage{amsmath}
\usepackage{wrapfig}
\usepackage{nicefrac}
\usepackage{cite}
\usepackage{bm}
\usepackage[font=footnotesize,skip=10pt]{caption}

\def\Journal#1#2#3#4{{#1} {\bf #2}, #3 (#4)}

\def\PRD{{\em Phys. Rev.} D}

\def\be{\begin{equation}}
\def\ee{\end{equation}}
\def\bea{\begin{eqnarray}}
\def\eea{\end{eqnarray}}

\newcommand{\Photo}{}

\begin{document}
\vspace*{4cm}
\title{The Simons Observatory: Combining delensing and foreground cleaning \\ for improved constraints on inflation}

\author{Emilie Hertig$^{1}$ \footnote{emh83@cam.ac.uk}, Kevin Wolz$^2$, Toshiya Namikawa$^3$, Antón Baleato Lizancos$^4$, Susanna Azzoni$^5$ and Anthony Challinor$^{1,6}$, on behalf of the Simons Observatory Collaboration}

\address{$^1$IoA and KICC, University of Cambridge, Madingley Road, Cambridge, CB3 0HA, UK \\ $^2$Department of Physics, University of Oxford, Keble Road, Oxford, OX1 3RH, UK \\ $^3$CD3, Kavli IPMU (WPI),
UTIAS, The University of Tokyo, Kashiwa, 277-8583, Japan \\ $^4$Berkeley Center for Cosmological Physics, UC Berkeley, CA 94720, USA \\ $^5$Department of Astrophysical Sciences, Peyton Hall, Princeton University, Princeton, NJ 08544,
USA \\ $^6$DAMTP, Centre for Mathematical Sciences, Wilberforce Road, Cambridge, CB3 0WA, UK}

\maketitle
\abstracts{The Simons Observatory (SO), a next-generation ground-based CMB experiment in its final stages of construction, will target primordial $B$-modes with unprecedented sensitivity to set tight bounds on the amplitude of inflationary gravitational waves. Aiming to infer the tensor-to-scalar ratio $r$ with precision $\sigma(r=0)\leq0.003$, SO will rely on powerful component-separation algorithms to distinguish the faint primordial signal from stronger sources of large-scale $B$-modes such as Galactic foregrounds and weak gravitational lensing. We present an analysis pipeline that performs delensing and foreground cleaning simultaneously by including multifrequency CMB data and a lensing $B$-mode template in a power-spectrum-based likelihood. Here, we demonstrate this algorithm on masked SO-like simulations containing inhomogeneous noise and non-Gaussian foregrounds. The lensing convergence is reconstructed from high-resolution simulations of the CMB and external mass tracers. Using optimized pixel weights for power spectrum estimation, the target precision for SO's nominal design is achieved and delensing reduces $\sigma(r)$ by 27--37\%, depending on foreground complexity.}

\section{Introduction}

Polarization patterns in the cosmic microwave background (CMB) encode essential information on the fundamental physics of the very early universe. In particular, detecting the $B$-mode signature of primordial gravitational waves (PGW) would provide compelling evidence in favour of cosmic inflation and tightly constrain the processes driving it. Degree-scale $B$-modes are therefore a primary target of next-generation CMB experiments such as the Simons Observatory (SO). The amplitude of PGW is parametrized by the tensor-to-scalar ratio $r$; SO's nominal design including three small-aperture telescopes (SATs) will aim to detect or rule out $r > 0.01$ at the $3\sigma$ level~\cite{SO}, thus significantly improving upon the current upper bound $r < 0.032$~\cite{tristram}.

Secondary large-scale $B$-modes sourced by the weak gravitational lensing of CMB photons~\cite{lensing} are becoming an important limitation for such sensitive instruments: the lensing $B$-mode contribution $\sigma_{\rm{lens}}^2$ to sample variance on $r$ is comparable to $(5$ $\mu$K-arcmin$)^2$ white noise, and therefore surpasses SO's goal level of $2$ $\mu$K-arcmin. This effect is mitigated by building a map of the lensing $B$-modes using high-resolution CMB data from the large-aperture telescope (LAT) and external large-scale structure (LSS) tracers (see Sec.~\ref{sec:theory}). Polarized Galactic foregrounds, in particular thermal dust emission and synchrotron radiation, are another significant contaminant. By gathering data in six frequency bands, SO will allow us to disentangle these components based on their characteristic spectral energy distributions (SEDs) (see Sec.~\ref{sec:pipeline}). 

In the present work (discussed in more detail in~\cite{hertig}), a parametric foreground model is fitted at the power spectrum level~\cite{wolz} and delensing is performed by including auto- and cross-spectra between the lensing template and the SAT $B$-modes in the likelihood. Such delensing was demonstrated for SO on foreground-free simulations in~\cite{namikawa}. A similar approach was first applied to BICEP/Keck data~\cite{bicep_del}, leading to a 10\% improvement in $\sigma(r)$ after delensing.

\section{Theoretical framework}\label{sec:theory}

The deflection of CMB photons by LSS along the line of sight results in a direction-dependent remapping of polarization anisotropies that mixes $E$- and $B$-modes. At leading order, the lensing-induced $B$-mode component is given by~\cite{namikawa}
\begin{equation}\label{eq:lensing_B}
    B^{\textrm{lens}}_{lm}=-i\sum_{l',m'}{\sum_{L,M}{\left(\begin{matrix} l & l'& L \\ m & m' & M \end{matrix}\right)p^{-}F^{(2)}_{ll'L}E^{*}_{l'm'}\kappa^{*}_{LM}}},
\end{equation}
where the factor $p^{-}$ is 0 or 1 for $l+l'+L$ even or odd, respectively, and $\kappa$ is the lensing convergence. Due to the mode-coupling encoded by $F^{(2)}_{ll'L}$, high-resolution $E$-modes are required to build an accurate template of the large-scale lensing $B$-modes. We extract these from our LAT-like simulations and apply a diagonal Wiener filter to mitigate instrumental noise, thus obtaining $\hat{E}_{lm}^{\rm{WF}}=[C_l^{EE} /\left(C_l^{EE}+N_l^{EE}\right)]\hat{E}_{lm} $, where $C_l^{EE}$ and $N_l^{EE}$ are the signal and noise power spectra, respectively.

Still using LAT-like mock data, $\kappa$ is first reconstructed internally through quadratic estimators (QEs)~\cite{okamoto}, which are derived from off-diagonal ($l \neq l'$) terms in the 2-point correlators of lensed CMB fields. QEs are computed for four field pairings (TT, TE, EE and EB) and combined to form a minimum-variance estimator. At $L>250$, QEs become dominated by reconstruction noise but may be complemented by external LSS tracers. Here, we use simulated Gaussian CIB and galaxy-overdensity maps informed by Planck data and LSST forecasts~\cite{yu}. Our final estimator is given by a weighted sum of all tracers $\hat{\kappa}^i$,
\begin{equation}\label{eq:kappa_comb}
    \hat{\kappa}^{\textrm{comb}}_{LM}=\sum_{ij}(\bm{\rho}^{-1}_{L})^{ij}\rho^{j\kappa}_{L}\sqrt{\frac{C_L^{\kappa\kappa}}{C_L^{\hat{\kappa}^{i}\hat{\kappa}^{i}}}}\hat{\kappa}^{i}_{LM},
\end{equation}
chosen to maximize the correlation between $\hat{\kappa}^{\textrm{comb}}$ and the true lensing convergence $\kappa$ (where $\bm{\rho}_L$ represents the matrix of correlation coefficients for our set of tracers~\cite{yu} whose power spectra are $C_L^{\hat{\kappa}^{i}\hat{\kappa}^{i}}$).

The ensemble-averaged power spectrum of the lensing template $B_t$, obtained by substituting $\hat{E}_{lm}^{\rm{WF}}$ and $\hat{\kappa}_{LM}^{\textrm{comb}}$ into Eq.~\eqref{eq:lensing_B}, can be shown to be identical to its cross-spectrum with the observed SAT $B$-modes~\cite{hertig}:
\begin{equation}
    C^{B_{t}B}_l = C^{B_{t}B_{t}}_l = \sum_{l'L}{|\mathcal{M}(l,l',L)|^2 \frac{C^{EE}_{l'}}{C^{\hat{E}\hat{E}}_{l'}}C^{EE}_{l'}C^{\kappa\hat{\kappa}^{\textrm{comb}}}_{L}}, 
    \label{eq:Cl_Bxtemp}
\end{equation}
where $\mathcal{M}(l,l',L)$ incorporates the prefactors of Eq.~\eqref{eq:lensing_B}. The Fisher information 
for $r$ then reduces to the same expression whether the template is subtracted at the map level or treated as an additional input to the likelihood. While the argument in~\cite{hertig} assumes a full-sky survey, it is expected to hold on the masked sky provided the same pixel weights are applied to all maps.

\section{Pipeline and simulations}\label{sec:pipeline}

All SO-like simulations used to test our pipeline assume we can meet the white-noise goals of SO's nominal design with three SATs~\cite{SO}. Inhomogeneous white and $1/f$ noise maps are generated based on the latest versions of the scanning strategy, and added to beam-convolved realizations of the lensed CMB. We neglect foregrounds for the LAT, but include dust and synchrotron emission in the SAT-like maps. These are modelled in \texttt{PySM3} with three levels of complexity: \texttt{d0s0} assuming uniform spectral properties, as well as \texttt{d1s1} and \texttt{dmsm} including realistic SED spatial variations.

\begin{wrapfigure}{l}{0.48\textwidth}
    \vspace{-10pt}
    \centering
    \includegraphics[width=\linewidth, height=0.29\textheight]{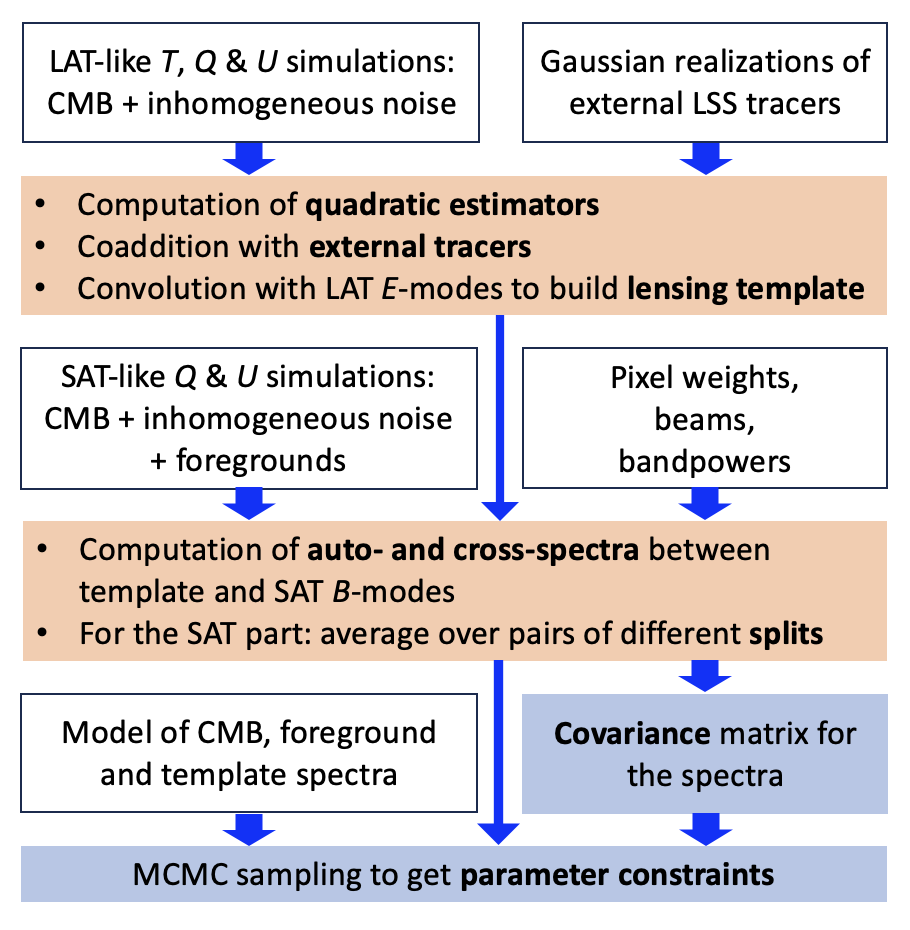}
    \caption[width=0.5\linewidth]{Flowchart of our pipeline. Operations in orange boxes are carried out on both the mock data and the fiducial simulations, while the blue-colored stages are only applied to the mock data.}
    \label{fig:flowchart}
    \vspace{-10pt}
\end{wrapfigure}

Using \texttt{NaMaster}, we construct mask-deconvolved and purified pseudo-$C_l$ estimates of the 28 auto- and cross-spectra between the SAT-like maps and the lensing $B$-mode template $B_t$. We apply local pixel weights 
\begin{equation}\label{eq:weights}
w_i \propto \left(A_{\rm{lens}}\sigma_{\textrm{lens}}^2+\sigma_{\textrm{white}}^2/N_i^{\textrm{hit}}\right)^{-1},
\end{equation}
where $A_{\rm{lens}}=1-\overline{C_l^{B_tB_t}/C_l^{BB}}$ represents the fractional residual lensing $B$-mode power averaged over $l$, and the central white-noise variance $\sigma_{\rm{white}}^2$ is modulated by the hit counts in pixel $i$.

In our parametric model for the CMB, $rC_{l}^{\textrm{tens}}+C_{l}^{\textrm{lens}}$, the lensing component is kept fixed as it is very well constrained within a given cosmology. Following~\cite{wolz}, we model foreground power spectra as $C_l^c=A_c(l/80)^{\alpha_c}$ with $c\in\{d,s\}$ for dust and synchrotron; their SEDs are respectively described by a modified blackbody function and a power law, with spectral indices $\beta_d$ and $\beta_s$ and a fixed temperature $T_d=20$ K. For input maps with spatially varying foreground SEDs, we mitigate potential biases by performing the moment expansion described in~\cite{azzoni}. Fluctuations in $\beta_c$ ($c\in\{d,s\}$) are then encoded by additional power spectra $C_l^{\beta_c}=B_c(l/80)^{\gamma_c}$.

Posterior distributions for all parameters are obtained by sampling a Gaussian likelihood for the power spectra, using a fiducial covariance matrix precomputed from 500 realizations of the CMB, SAT-like noise and Gaussian foregrounds.

\section{Results}

\begin{figure}[h]
    \centering
    \includegraphics[width=\linewidth,height=0.17\textheight]{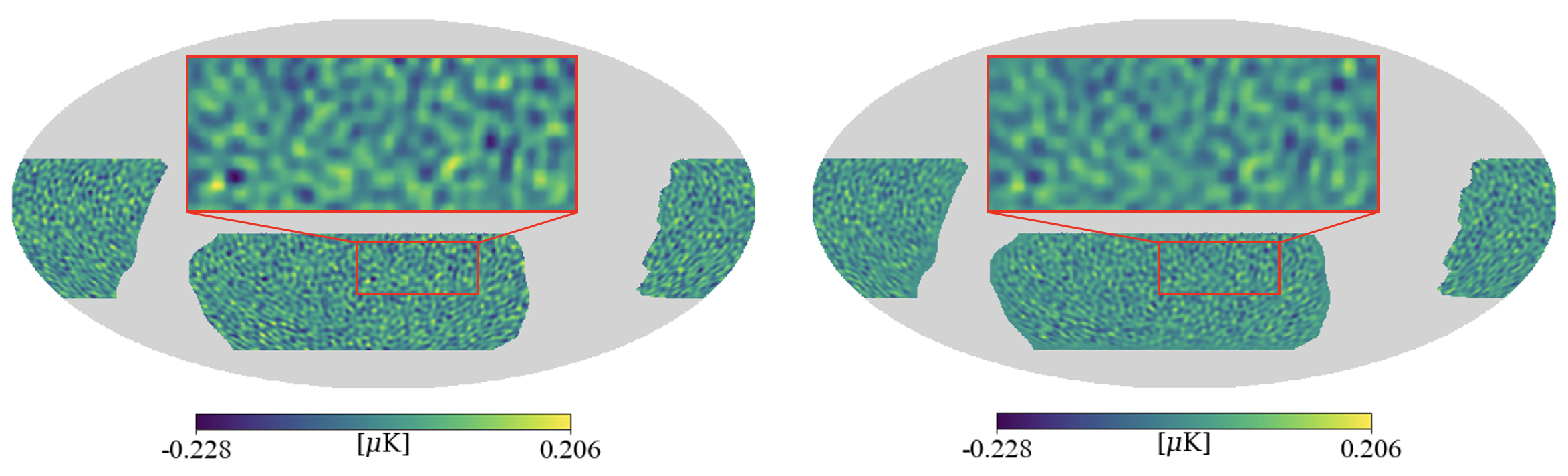}
    \caption[]{\textit{Left}: lensing $B$-modes in one of the input CMB maps. \textit{Right}: corresponding lensing $B$-mode template. Both are projected onto the overlap area between the SAT and LAT masks, keeping multipoles $20 \leq l \leq 128$.}
    \label{fig:template_maps}
    \vspace{-5pt}
\end{figure}

As shown in Fig.~\ref{fig:template_maps}, our lensing templates are clearly correlated with the $B$-modes in the corresponding input CMB simulations. The Wiener filters applied to $\hat{E}^{\rm{WF}}$ and $\hat{\kappa}^{\rm{comb}}$ downweight noisy multipoles in Eq.~\eqref{eq:lensing_B}, resulting in a lower amplitude for the reconstructed lensing $B$-modes; with our simulations, we achieve a delensing efficiency of 65\% ($A_{\rm{lens}}=0.35$).

\begin{figure}[h!]
    \centering
    \begin{minipage}{.47\textwidth}
        \centering
        \includegraphics[width=\linewidth,height=0.17\textheight]{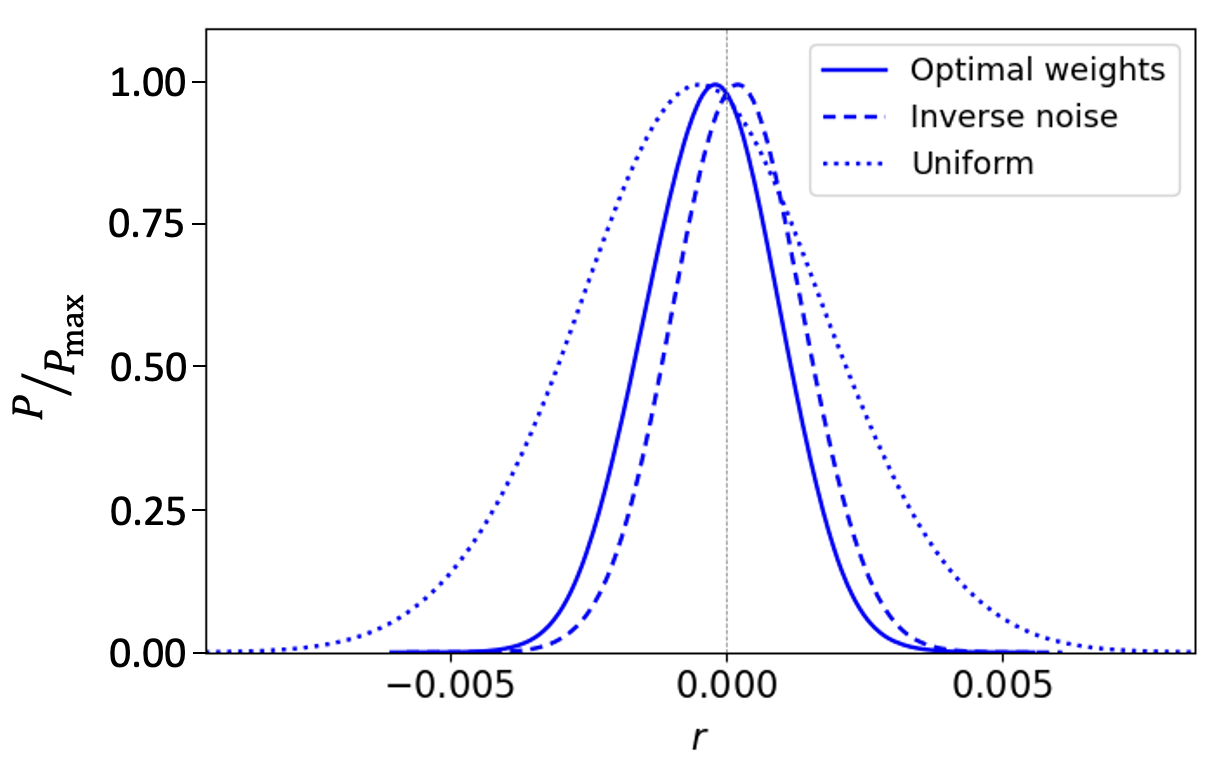}
        \caption[width=\linewidth]{Marginalized posterior distributions for $r$ with delensing, using a single input map with \texttt{d0s0} foregrounds and three different pixel weights.}
        \label{fig:mask_comp}
    \end{minipage}
    \hfill
    \begin{minipage}{.48\textwidth}
        \centering
        \includegraphics[width=\linewidth,height=0.17\textheight]{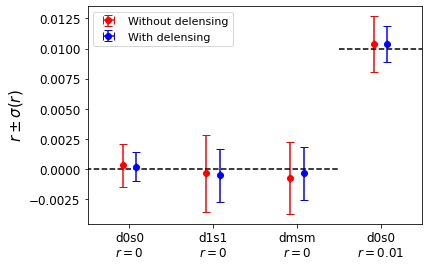}
        \caption[width=\linewidth]{Best-fit $\hat{r}$ and MCMC standard deviation $\sigma(r)$ with and without delensing, averaged over 100 simulations for each foreground and CMB model.}
        \label{fig:del_performance}
    \end{minipage}
\vspace{-10pt}
\end{figure}

Figure~\ref{fig:mask_comp} compares constraints on $r$ obtained using the optimized pixel weights in Eq.~\eqref{eq:weights} to those obtained with other weighting schemes used in previous publications. While an apodized uniform mask is clearly suboptimal due to the high effective noise level in the outer regions of the survey area, the hit-count mask proportional to the inverse noise variance performs very similarly to our method. This is expected at SO's nominal sensitivity, as instrumental noise still dominates over lensing residuals with $A_{\rm{lens}}=0.35$. However, Eq.~\eqref{eq:weights} will become increasingly relevant for future experiments with lower noise levels, for which the lensing residuals will be comparatively more important and may vary across the sky.

The summary statistics for 100 realizations displayed in Fig.~\ref{fig:del_performance} verify that our implementation of delensing does not bias the best-fit estimates $\hat{r}$ and that non-zero tensor modes are successfully detected by the pipeline. In the absence of PGW, delensing reduces uncertainties on $r$ by 37\%, 31\% and 27\% for \texttt{d0s0}, \texttt{d1s1} and \texttt{dmsm} foregrounds, respectively. In the latter two cases with spatially varying SEDs, the larger error bars result from sampling four additional parameters when applying the moment-expansion method. As expected, delensed $\sigma(r)$ values are consistent with those obtained for input simulations containing 65\% less lensing $B$-mode power~\cite{hertig}.

\section{Outlook}

Preparing our pipeline for application to early SO data will require characterizing the impact of LAT foreground residuals in the lensing template and accounting for SAT timestream filtering. Beyond this, the present work will become increasingly relevant after the planned addition of three SO SATs, as well as for upcoming Stage-4 experiments: at such sensitivity levels, efficient delensing will be crucial in further reducing uncertainties on $r$.


\section*{Acknowledgments}

This work was supported in part by a grant from the Simons Foundation (Award \#457687, B.K.). EH is supported by a Gates Cambridge Scholarship (grant OPP1144).


\bibliography{moriond}

\end{document}